\documentstyle[aps]{revtex}
\begin{document}
\input{psfig}
\draft
\title{\bf Long-time-scale revivals in ion traps}
\author{H. Moya-Cessa\footnote{Electronic address: hmmc@inaoep.mx}
\thanks{Permanent address:INAOE, Coordinaci\'on de Optica, Apdo. Postal 51 y 216, 72000 
Puebla, Pue., Mexico}, 
A. Vidiella-Barranco\footnote{Electronic address: vidiella@ifi.unicamp.br}, 
J.A. Roversi\footnote{Electronic address: roversi@ifi.unicamp.br}, 
Dagoberto S. Freitas\footnote{Electronic address: dfreitas@ifi.unicamp.br}
\thanks{Permanent address:
Universidade Estadual de Feira de Santana, Departamento de F\'\i sica, 
44031-460, Feira de Santana, BA, Brazil},\\ 
and S.M. Dutra\footnote{Electronic address: dutra@rulhm1.leidenuniv.nl}}
\address{Instituto de F\'\i sica ``Gleb Wataghin'',
Universidade Estadual de Campinas,
13083-970   Campinas  SP  Brazil}
\date{\today}
\maketitle
\begin{abstract}
In this contribution we investigate the interaction of a single 
ion in a trap with laser beams. 
Our approach, based on unitary transformating the Hamiltonian,
allows its exact diagonalization without performing the Lamb-Dicke 
approximation. We obtain a transformed Jaynes-Cummings type Hamiltonian,  
and we demonstrate the existence of super-revivals in that system. 
\end{abstract}

There has been recently a great deal of interest in single trapped ions interacting
with laser beams. This system allows the preparation of nonclassical states of the
vibrational motion of the ion \cite{cira1,vogel}. In fact, the generation of Fock, coherent, 
squeezed, \cite{meek} and Schr\"odinger-cat states \cite{monr} has been already accomplished. 
Besides, there are potential practical possibilities in several fields, such as precision 
spectroscopy \cite{wine} and quantum computation \cite{cira0}, for instance.

We would like to remark that the theoretical treatment of the interaction of a trapped 
ion with one or several laser beams constitutes a complicated problem.
Normally the problem is solved in limiting situations, e.g., in the Lamb-Dicke
regime, in which the ion is confined in a region much smaller than the laser wavelenght.
Other limiting situations concern the laser intensity; if the effective Rabi 
frequency $\Omega$ of the ion-laser interaction is such that $\Omega\ll \nu$, where 
$\nu$ is the trap frequency \cite{bloc,cira2}, we have the low-excitation
regime. Otherwise, if  $\Omega\gg \nu$, this corresponds to the strong-excitation regime  
\cite{poya}. 

Here we adopt a new approach to this problem. We depart from the full ion-laser
Hamiltonian, and perform a unitary transformation that allows us to obtain a 
Jaynes-Cummings like Hamiltonian without the rotating-wave approximation (RWA). We therefore
are able to analitically solve the problem, by applying the RWA, which corresponds to the 
regime $\Omega\approx\nu$. We should stress that our method makes possible to close 
the gap between the different intensity regimes, and moreover, with no restrictions on 
the Lamb-Dicke parameter. 

We consider a single trapped ion interacting with two laser plane waves 
(frequencies $\omega_1$ and $\omega_2$), in a Raman-type configuration. 
Both laser beams, as usual, will be treated as classical fields propagating along 
the $x$ axis, so that we have a one-dimensional problem. The corresponding scheme of
levels is shown in Fig. (1). The lasers effectively drive the electric-dipole forbidden 
transition $|g\rangle\leftrightarrow|e\rangle$ (frequency $\omega_0$), and we may have 
a detuning $\delta=\omega_0-\omega_L$, where ($\omega_L=\omega_1-\omega_2$). For a
sufficiently large detuning $\Delta$, the third level $|r\rangle$ may be adiabatically
eliminated \cite{gerr}, and we end up with an effective two-level system.
This situation is described by the following Hamiltonian \cite{poya}
\begin{equation}
\hat{H}= \hbar\nu \hat{a}^{\dagger}\hat{a} + \hbar\frac{\delta}{2}\sigma_z +
\hbar\frac{\Omega}{2}\left( \hat{\sigma}_{-} 
e^{-i\eta(\hat{a}+\hat{a}^\dagger)} +\hat{\sigma}_{+} 
e^{i\eta(\hat{a}+\hat{a}^\dagger)} \right),
\label{H}
\end{equation}
being $\eta$ the Lamb-Dicke parameter, 
$\hat{\sigma}_{+}=|e\rangle\langle g|$, ($\hat{\sigma}_{-}=|g\rangle\langle e|$)
the usual electronic raising (lowering) operator, and $\hat{a}^{\dagger}$ 
($\hat{a}$) the ion's vibrational creation (annihilation) operator. 

By applying the unitary transformation
\begin{equation}
\hat{T}
= {1\over \sqrt{2}} 
\left\{ 
{1\over 2} \left[
\hat{D}^{\dagger}\left(\beta \right)
+ \hat{D}\left(\beta \right) \right]\hat{I}
+ {1\over 2} \left[
\hat{D}^{\dagger}\left(\beta\right)
-\hat{D}\left(\beta \right)
\right] \hat{\sigma}_z
+\hat{D}\left(\beta \right)\hat{\sigma}_{+}
-\hat{D}^{\dagger}\left(\beta \right) \hat{\sigma}_{-}
\right\},\label{T}
\end{equation}
to the Hamiltonian in Eq. (\ref{H}), where $\hat{D}(\beta)=\exp(\beta\hat{a}^\dagger-
\beta^*\hat{a})$ is Glauber's displacement
operator, with $\beta=i\eta/2$, we obtain the following transformed 
Hamiltonian 
\begin{equation}
\hat{\cal H}
\equiv \hat{T}\hat{H}\hat{T}^{\dagger}  
= \hbar \nu \hat{n}  
+ \hbar\Omega\hat{\sigma}_z -i\hbar{\eta\nu\over 2}  
\left[ (\hat{a}^{\dagger}-\hat{a}) -i\frac{\delta}{\eta\nu}\right] 
\left(\hat{\sigma}_{-}+\hat{\sigma}_{+}\right)+ \hbar \nu {\eta^2 \over 4}.
\end{equation}
This result holds for any value of the Lamb-Dicke parameter $\eta$. Our Hamiltonian 
in Eq. (\ref{H}) becomes a Jaynes-Cummings-type Hamiltonian, and therefore its   
diagonalization is allowed provided we perform the rotating wave approximation (RWA). 
This is possible in the regime $\nu=2\Omega$, i.e.,
when the Rabi frequency of the coupling between the laser and the ion ($\Omega$), 
is of the order of the vibrational frequency of the ion in the trap ($\nu$).

Rewriting $\hat{\cal H}$ in the interaction representation, 
\begin{equation}
\hat{\cal H}
= -i\hbar{\eta\nu\over 2}  
\left[\left(\hat{a}^{\dagger}\hat{\sigma}_{-}e^{i(\nu-2\Omega)t}-h.c.\right)+
\left(\hat{a}^{\dagger}\hat{\sigma}_{+}e^{i(\nu+2\Omega)t}-h.c.\right)+
-i\frac{\delta}{\eta\nu}
\left(\hat{\sigma}_{+}e^{2i\Omega t}-h.c.\right)\right],\label{THT}
\end{equation}
we see that we may neglect the last two rapidly oscillating terms in the right hand 
side of Eq. (\ref{THT}), which corresponds to the RWA. We then obtain

\begin{equation}
\hat{\cal H}\approx \hat{\cal H}_{\scriptscriptstyle{RWA}}=\hbar \nu \hat{n}  
+ \hbar\frac{\nu}{2}\hat{\sigma}_z -i\hbar g  
\left( \hat{a}^{\dagger}\hat{\sigma}_{-}-\hat{a}\hat{\sigma}_{+}\right),
\label{JCH}
\end{equation}
which coincides with the Jaynes-Cummings Hamiltonian. The effective coupling constant
is $g =\eta\nu/2$, and the term $\hbar \nu \eta^2/ 4$ has not been taken into account 
because it just represents an overall phase. We note that our final result in Eq. (\ref{JCH})
is valid even for non-zero detunings $\delta$.

The time evolution of the state vector, for an initial state $|\psi(0)\rangle$ is
\begin{equation}
|\psi(t) \rangle=\hat{T}^\dagger\hat{U}_I(t)\hat{T}|\psi(0) \rangle,
\end{equation}
where $\hat{U}_I(t)$ is the Jaynes-Cummings evolution operator \cite{sten} in the
interaction picture 
\begin{equation}
\hat{U}_I(t)={1\over \sqrt{2}} 
\left\{ 
{1\over 2} \left[\hat{C}_{n+1}+\hat{C}_{n}\right]\hat{I}
+ {1\over 2} \left[\hat{C}_{n+1}-\hat{C}_{n}\right] \hat{\sigma}_z
+\hat{S}_{n+1}\hat{a}\hat{\sigma}_{+}
-\hat{a}^\dagger\hat{S}_{n+1}\hat{\sigma}_{-}\right\},\label{U}
\end{equation}
with $\hat{C}_{n+1}=\cos\left(g t\sqrt{\hat{a}\hat{a}^\dagger}\right)$ and
$\hat{S}_{n+1}=\sin\left(g t\sqrt{\hat{a}\hat{a}^\dagger}\right)
/{\sqrt{\hat{a}\hat{a}^\dagger}}$.

For an initial state $|\psi(0) \rangle=|e\rangle|\alpha\rangle$, or
the atom in the excited state [see Fig. (1)] and the ion in a coherent (vibrational)
state, the mean number of vibrational quanta as a function of the (dimensionless)
scaled time $\tau=g t$, or $\langle \hat{n} \rangle=\langle\psi(\tau)|\hat{n}|\psi(\tau)\rangle$
is given by

\begin{eqnarray}
\langle \hat{n} \rangle & = & \frac{1}{2} \sum_{n=0}^{\infty} \left( \left| 
c_{n,e}(t) \right|^{2} + \left| c_{n,g}(t) \right|^{2} \right) \left( n + 
\frac{\eta^2}{4} \right)  \nonumber \\
  &  & +\left(\frac{i \eta }{4}\right) \sum_{n=0}^{\infty} \sqrt{n + 1} \left[
  c_{n+1,e}^{*}(t) c_{n,g}(t) - c_{n,e}^{*}(t) c_{n+1,g}(t) - c.c.\right]\label{nm}
\end{eqnarray}

where

\begin{equation}
c_{n,e}(t) = \exp{-\left|\tilde{\alpha}\right|^{2}/2} \left[ \frac{\tilde{\alpha}^n}
{\sqrt{n!}} \cos\left( gt \sqrt{n+1} \right) + 
\frac{\tilde{\alpha}^{n+1}}{\sqrt{(n+1)!}} \sin\left( gt 
\sqrt{n+1} \right) \right],
\end{equation}

\begin{equation}
c_{n,g}(t) = \exp{-\left|\tilde{\alpha}\right|^{2}/2} \left[ \frac{\tilde{\alpha}^{n-1}}
{\sqrt{(n-1)!}} \sin\left( gt \sqrt{n} \right) - 
\frac{\tilde{\alpha}^{n}}{\sqrt{n!}} \cos\left( gt \sqrt{n}\right) 
\right],
\end{equation}
and $\tilde{\alpha}=\alpha-i\eta/2$.

The structure of the equation above is similar to the one obtained 
for $\langle \hat{n} \rangle$ in the driven-Jaynes-Cummings model (DJCM) \cite{hmcsmd}. 
Therefore we expect the phenomenon of super-revivals to be present in the ion system, 
as an analogy to the DJCM. Super-revivals are long time scale revivals arising in the
atom-field dynamics. This peculiar behavior is illustrated in Fig. (2). The
super-revivals in $\langle \hat{n} \rangle$ occur for a time $\tau_{sr}\approx 4|\alpha|^2
\tau_r$ ($\tau_r=2\pi\sqrt{\overline{n}}/\Omega$ is the revival time). We note that the
existence or not of super-revivals is narrowly connected to the preparation of the initial
vibrational state. For instance, if we have $\alpha=(5.0,0.5)$ and $\eta=0.5$, 
the super-revivals do not occur [see Fig. (2a)], and we have ordinary revivals only. 
However, for appropriate values of the parameters $\alpha=(0.5,5.0)$ and $\eta=0.5$, 
super-revivals take place in that system [see Fig. (2b)].

We also would like to point out that because dissipation in ion traps is much smaller 
than in cavities, it might be allowed the experimental observation of super-revivals 
in that system. 

In conclusion, we have presented a novel approach for studying the dynamics of the
trapped ions driven by laser beams. We have succeeded in linearizing the Hamiltonian 
through the application of a unitary transformation. Then we were able to treat the 
problem in a specific intensity regime ($\Omega\approx\nu$), without the need of performing 
the Lamb-Dicke approximation. As a result we have predicted the existence of long 
time-scale revivals (super-revivals) for the excitation number in trapped ions.

\begin{figure}[hp]
\vspace{1cm}
\centerline{\hspace{1.0cm}\psfig{figure=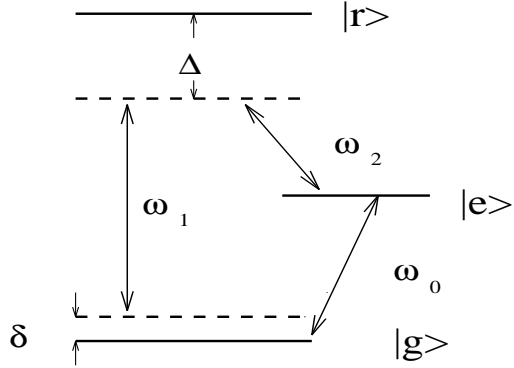,height=5cm,width=7cm}}
\vspace{1cm}
\caption{Configuration of levels of the trapped ion interacting with two laser
beams, of frequencies $\omega_1$ and $\omega_2$.}
\end{figure}

\begin{figure}[hp]
\vspace{1cm}
\centerline{\hspace{1.0cm}\psfig{figure=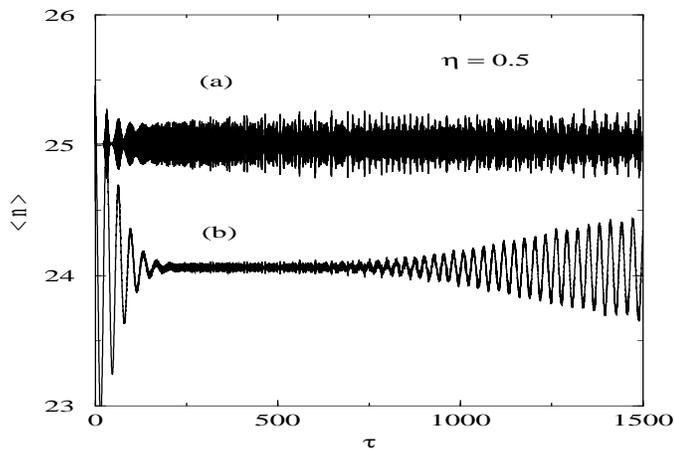,height=5cm,width=7cm}}
\vspace{1cm}
\caption{Plot of the mean excitation number $\langle\hat{a}^\dagger\hat{a}\rangle$
as a function of the (dimensionless) scaled time $\tau=g t$. 
a) Ordinary revivals occurring for $\alpha=(5.0,0.5)$, 
and b) super-revivals ocurring for $\alpha=(0.5,5.0)$. In both cases the Lamb-Dicke 
parameter is $\eta=0.5$.}
\end{figure}

\newpage 

One of us, H.M.-C., thanks W. Vogel for useful comments.
This work was partially supported by Consejo Nacional de
Ciencia y Tecnolog\'\i a (CONACyT), M\'exico, Conselho Nacional de 
Desenvolvimento Cient\'\i fico e Tecnol\'ogico (CNPq), Coordena\c c\~ao de
Aperfei\c coamento de Pessoal de N\'\i vel Superior (CAPES), Brazil, 
Funda\c c\~ao de Amparo \`a Pesquisa do Estado de S\~ao Paulo (FAPESP ), Brazil, 
and International Centre for Theoretical Physics (ICTP), Italy.

%
%

%
%

\end{document}